\documentclass[preprint,5p,times,twocolumn]{elsarticle}

\usepackage{amssymb}
\usepackage{amsmath}
\usepackage{graphicx}
\usepackage{caption}
\usepackage{subcaption}
\usepackage{xcolor}
\usepackage{algorithm}
\usepackage{algpseudocode}

\begin{document}

\begin{frontmatter}

\title{TopoEdit: Fast Post-Optimization Editing of Topology Optimized Structures}

\author{Hongrui Chen, Josephine V. Carstensen, Faez Ahmed} 

\affiliation{organization={Massachusetts Institute of Technology},
            city={Cambridge},
            state={MA},
            postcode={02139}
}

\begin{abstract}
Despite topology optimization producing high-performance structures, late-stage localized revisions remain brittle: direct density-space edits (e.g., warping pixels, inserting holes, swapping infill) can sever load paths and sharply degrade compliance, while re-running optimization is slow and may drift toward a qualitatively different design. We present TopoEdit, a fast post-optimization editor that demonstrates how structured latent embeddings from a pre-trained topology foundation model (OAT) can be repurposed as an interface for physics-aware engineering edits. Given an optimized topology, TopoEdit encodes it into OAT’s spatial latent, applies partial noising to preserve instance identity while increasing editability, and injects user intent through an edit-then-denoise diffusion pipeline. We instantiate three edit operators: drag-based topology warping with boundary-condition–consistent conditioning updates, shell–infill lattice replacement using a lattice-anchored reference latent with updated volume-fraction conditioning, and late-stage no-design region enforcement via masked latent overwrite followed by diffusion-based recovery. A consistency-preserving guided DDIM procedure localizes changes while allowing global structural adaptation; multiple candidates can be sampled and selected using a compliance-aware criterion, with optional short SIMP refinement for warps. 
Across diverse case studies and large edit sweeps, TopoEdit produces intention-aligned modifications that better preserve mechanical performance and avoid catastrophic failure modes compared to direct density-space edits, while generating edited candidates in sub-second diffusion time per sample.

\end{abstract}


\end{frontmatter}

\section{Introduction}
Topology optimization has become a powerful approach in engineering design, allowing the development of lightweight, high-performance structures through the optimal distribution of material within a specified design domain. However, the complex shape of the solutions generated by topology optimization often poses significant challenges for practical design workflows, where engineers must interpret, trust, and revise the optimized result to satisfy downstream considerations and design intent. In these settings, even small local changes, such as adjusting a joint location, replacing an interior region with an infill, or enforcing a last-minute no-design region, typically require re-running optimization, which is time-consuming and can drift toward a qualitatively different design. Localized editing offers a promising alternative: rather than restarting the full solver or regenerating a new sample, a designer should be able to “drag” structural features toward a target configuration while still leveraging the model’s learned physical priors to maintain mechanical performance. Despite this potential, there is a lack of methods for topology editing that support fast, localized, and physics-aware modifications of existing topology-optimized designs.

Current approaches that target user control in topology optimization can be broadly categorized into those that rely on restrictive problem reformulations, post-processing techniques, or human-in-the-loop optimization schemes. Methods that modify the optimization formulation, such as adding constraints or tuning feature-size/length-scale controls, can incorporate certain preferences, but they still require iterative solves and careful parameter choices, limiting responsiveness in topology editing settings \cite{Chandrasekhar2021Fourier}. Similarly, post-processing techniques can modify the geometry after optimization, but they are typically not context-aware and do not explicitly account for structural performance under the applied loads and boundary conditions during the edit \cite{Hu2015SupportSlimming}. Interactive topology optimization frameworks that incorporate user intent can fill a practical gap, yet they primarily steer ongoing optimization via region-of-interest feature-size changes or pattern guidance, rather than enabling targeted, direct edits to an already-optimized design \cite{ha2023humaninformed,schiffer2023hitop2,schiffer2024interactiveinfill}. Data-driven generative models accelerate topology generation and can generalize across diverse conditions, but existing pipelines are typically designed to generate a solution from a specification rather than to edit a specific solution instance while preserving its identity \cite{nobari2024nito,nie2021topologygan}. In this work, we build upon these advances and describe a method that takes into account post-optimization editability and fast, localized control, while preserving physics-aware structural behavior.

A key to achieving our goal lies in the ability to perform localized edits without discarding the physical information encoded in the original optimized topology. Directly editing the density field in image space, such as warping pixels or inserting holes or infill, is ineffective. These operations can easily disrupt load paths or disconnect critical boundary-condition attachments, leading to significant compliance degradation (for example, direct warp in Figure \ref{fig:warp_example} (b)). Moreover, treating editing as iterative optimization is often slow, as it requires the user to intervene during the optimization and then still perform many small update steps to achieve the intended semantic change \cite{schiffer2023hitop2}. Additionally, enforcing practical edit constraints, such as preserving similarity, respecting boundary conditions near the edit region, and maintaining volume-fraction consistency, these constraints and objectives further complicate successful editing.

\noindent Our main contributions are:

\begin{itemize}
\item TopoEdit: We show that the structured latent embeddings of a pre-trained topology foundation model (OAT latent diffusion) can be repurposed for engineering post-optimization edits, enabling fast, instance-preserving modifications via partial noising and an edit-then-denoise pipeline.

\item Unified latent edit operators: We introduce three practical edit operators---topology warping, lattice infill replacement, and no-design region enforcement, implemented as latent interventions with edit-specific conditioning updates.

\item Physics-aware guidance and refinement: We propose a consistency-preserving guided DDIM denoising procedure (with localized reference guidance for warps), plus compliance-aware best-of-N candidate selection, and an optional short SIMP post-processing step to refine edited results.

\item Empirical validation of edit quality and efficiency: Through case studies and quantitative evaluation, we show TopoEdit achieves localized, intention-aligned edits while better preserving mechanical performance than direct density-space edits, with sub-second diffusion sampling in the reported setup.

\end{itemize}

\begin{figure*}
\centering

\includegraphics[width=\textwidth]{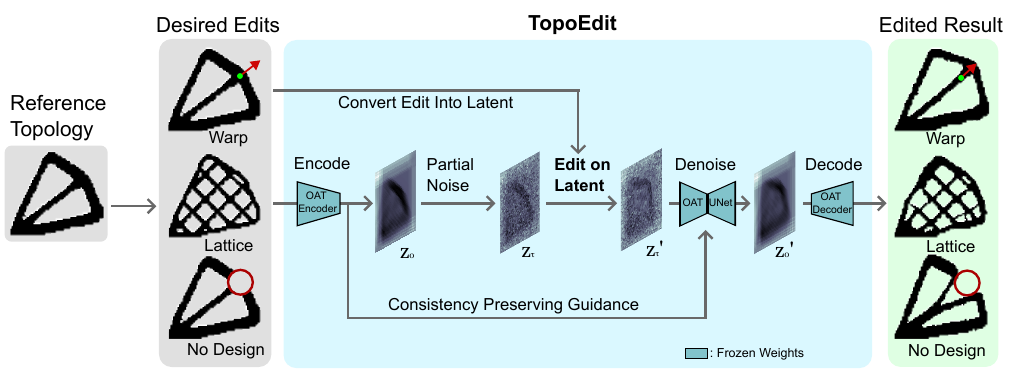}
\caption{TopoEdit uses a pretrained OAT with frozen weights. Given a reference topology and a user-specified desired edit (topology warp, lattice infill replacement, or no-design region enforcement), TopoEdit encodes the input into OAT’s structured latent, applies partial noising, and converts the edit into a corresponding latent-space intervention to produce an edited initialization. Starting from the edited latent, we run diffusion denoising with consistency-preserving guidance to recover an edited clean latent that preserves the instance identity away from the edit while allowing localized structural adaptation. Finally, the edited latent is decoded to yield the edited topology.}

\label{fig:flowchart}
\end{figure*}

\section{Previous Work}
Our review focuses on data-driven topology optimization, editing in topology optimization, and diffusion models. 

\paragraph{Data-driven topology optimization} Topology optimization distributes material within a design domain to satisfy performance objectives under constraints such as volume fraction. Early works include homogenization-based formulations \cite{bendsoe1988homogenization}, shape/topology perspectives beyond homogenization \cite{rozvany1992generalized}, Solid Isotropic Materials with Penalization (SIMP) interpolation \cite{bendsoe1999material}, and level-set approaches \cite{sethian2000structuralboundary,allaire2004sensitivitylevelset}. These methods remain widely used and supported by compact educational implementations \cite{sigmund2001top99}; however, their slow and iterative nature motivates faster surrogates for high-throughput exploration and interaction.

Data-driven methods skip optimization by learning a direct mapping from problem setup to a near-optimal topology. Machine learning methods include conditional GANs and transfer learning \cite{behzadi2022gantl,sharpe2019cgan}, supervised iteration-free predictors \cite{li2019noniterative,yu2019nearoptimal}, hybrids that couple learned components with optimization structure \cite{zhang2021tonr}, and architectures designed to improve generalization and extend beyond 2D settings \cite{wang2022perceptiblecnn,rawat2019conditional,zheng2021unet3d,keshavarzzadeh2021ddnsm}. More recent work also explores implicit representations to reduce fixed-resolution dependence \cite{hu2024iftonir}. While these approaches substantially accelerate generation, most are not designed for instance-preserving edits. Editing one optimized design rather than generating a new plausible design can be difficult with aformentioned data-driven method.

Generative modeling for topology progressed from GANs conditioned on physical fields \cite{nie2021topologygan} to resolution-free implicit generators \cite{nobari2024nito}, and most recently to diffusion-based models \cite{maze2023diffusionbeatsgans, giannone2023diffusing, giannone2023aligning} and finally foundation modeling (OAT) that targets shape- and resolution-free structural topology distributions \cite{nobari2025oat}. OAT is trained on 2.2 million optimized structures. These structures consist of 2 million unique boundary conditions, varying numbers of loads, and volume fractions. As a latent diffusion model, OAT significantly reduces the mean compliance by up to 90\% compared to the best prior models and achieves sub-second inference time. Additionally, OAT demonstrates excellent generalizability, performing well on out-of-distribution datasets like heatsink optimization. To edit topology-optimized results, the edit moves the result outside the original distribution of the trained manifold. This motivates us to leverage the model’s generalizability to generate edited topologies. The versatile latent representation of OAT aligns well with recent advancements in latent-based image editing using latent diffusion models. 

\paragraph{Editing in topology optimization}
Raw topology optimization results can meet structural requirements, but can be difficult to turn into a manufacturable design. Thus, a line of work focuses on making topology optimization editable by exposing handles for designer intent and localized control. One class of methods introduces explicit feature-size, thickness, and porosity controls to shape member widths and manufacturable detail \cite{ha2023humaninformed,schiffer2023hitop2,allaire2016thicknesscontrol,guo2014explicitfeature,schmidt2019gradedporosity}. Another class enables appearance-guided design, where users steer infill motifs or geometric textures through drawn patterns, pattern libraries, or appearance constraints, extending from 2D layouts to patterned shells and multi-pattern trade-offs \cite{schiffer2024interactiveinfill,navez2022patternlibrary,zhu2024holeappearance,meng2023freeformshells,li2024patternshellinfill,zhang2024multipattern}. In parallel, interactive frameworks incorporate subjective preferences and sketches as inputs to guide exploration and promote diverse candidate structures \cite{li2023subjective,li2025vrpreference,zhu2025sketchguided,zhang2024sketchaided,mueller2015designerprefs}. Beyond topology optimization specific formulations, patch-based correspondence and synthesis in graphics provides a conceptual toolkit for local replacement and pattern transfer \cite{barnes2009patchmatch,barnes2011patchmatch,barnes2017patchsurvey,khojasteh2020patchsynthesis}, and related ideas have been adapted to inject geometric priors into topology optimization via neural style transfer \cite{vulimiri2021neuralstyletopo}.

Despite this progress, many topology optimization editing mechanisms remain constrained by how they incorporate intent. Methods that encode edits as additional constraints or controls are typically physics-aware, but they still require reformulating and re-solving a global, iterative optimization, which is slow and can drift away from the original topology. Human-in-the-loop approaches similarly improve controllability, but they primarily steer an ongoing solve and do not directly support post-hoc edits of an already-optimized design without continued iterations. At the other extreme, post-processing geometry operations such as warps, hole insertion, or infill replacement can be fast, but they are not inherently aware of the boundary conditions; even localized changes may sever load paths or detach boundary-condition attachments, leading to large compliance degradation. Together, these limitations motivate an edit mechanism that is localized while remaining better aligned with the structural semantics of optimized designs.

\paragraph{Diffusion models and edits}
Diffusion models learn data distributions by inverting a noising process \cite{sohldickstein2015nonequilibrium,ho2020ddpm,song2021scorebasedsde}, with practical accelerations and control via implicit sampling and guidance \cite{song2021ddim,ho2022classifierfreeguidance} and strong empirical image-generation performance \cite{dhariwal2021diffusionbeatgans}. Latent diffusion improves efficiency by performing diffusion in a compressed latent space \cite{rombach2022latentdiffusion}, with few-step sampling further enabled by distillation and consistency formulations \cite{salimans2022progressivedistillation,song2023consistencymodels,luo2023latentconsistency}. Building on this, diffusion-based editing has been demonstrated for semantic/text-guided manipulation \cite{kim2022diffusionclip,hertz2022prompttoprompt,brooks2023instructpix2pix,kawar2023imagic,mokady2023nulltext,cao2023masactrl} and for interactive drag/point edits that propagate sparse constraints through learned features \cite{pan2023dragyourgan,shi2024dragdiffusion,mou2024dragondiffusion,mou2024diffeditor,ling2023freedrag,zhao2024fastdrag,liu2024dragyournose}. A consistent theme is that editing in a learned latent/feature space can preserve instance identity while enabling coherent local geometric change, often more robustly than direct pixel-space warping.

Motivated by latent editing in image-focused diffusion models, we investigate whether analogous principles can enable instance-preserving edits of topology-optimized designs. OAT is particularly suitable as a foundation because it models a broad distribution of optimized topologies through a latent diffusion pipeline \cite{nobari2025oat}, allowing edits to be expressed compactly in latent space while remaining compatible with diverse shapes and resolutions.

\section{Proposed Method}
Our method performs topology edits by manipulating the latent vectors of a diffusion model. We first encode the optimized density field into the latent space used by OAT (Section 3.1), then apply partial noising so the latent remains close to the original design manifold while introducing controllable stochasticity for editability. Given this partially noised state, we inject the user’s edit in an edit-then-denoise pipeline (Section 3.2). We then run guided denoising with a target latent that encodes the intended post-edit structure, using this reference to enforce consistency with the desired outcome while leveraging the UNet’s locality bias to keep edits predominantly local (Section 3.3), and finally decode the resulting latents back to density fields, with an optional short SIMP refinement step for the topology warp (Section 3.4). 

\subsection{Diffusion Model}
We build our topology editing pipeline on top of the pre-trained OAT latent diffusion model~\cite{nobari2025oat}. OAT is particularly well-suited for topology editing because it compresses a topology of arbitrary aspect ratio and resolution into a fixed, structured spatial latent and models the space of feasible topologies as a conditional probability distribution, enabling fast sampling of multiple post-edit candidates by noising and denoising in latent space. Here, we provide a brief overview of the architecture of OAT. 

\paragraph{Neural-field autoencoder}
Let $T \in [0,1]^{H\times W}$ denote an optimized density field on a rectangular design domain. OAT first maps $T$ into a fixed-resolution latent $z$ using a neural-field autoencoder (which we refer to as the NFAE):
\begin{equation}
z = E(T), \qquad \phi = D(z), \qquad \tilde{T} = R(\phi, c, s).
\end{equation}
Here, $E(\cdot)$ is an encoder, $D(\cdot)$ is a decoder producing a feature tensor $\phi$, and $R(\cdot)$ is a neural renderer that reconstructs the topology at arbitrary resolution using (i) spatial coordinates $c$ (pixel/element centers) and (ii) cell/pixel sizes $s$. In OAT, variable-resolution/shape inputs are padded and resized to a fixed resolution for encoding, while the renderer enables decoding back to the native (or any desired) resolution. The autoencoder is trained with an $\ell_1$ reconstruction objective,
\begin{equation}
\mathcal{L}_{\mathrm{AE}} = \| \tilde{T} - T_{\mathrm{GT}}\|_{1},
\end{equation}
where $T_{\mathrm{GT}}$ denotes the ground-truth topology and $\tilde{T}$ is its reconstruction.

\paragraph{Problem representation for conditional generation}
Each topology optimization instance is defined by the domain shape, loads, fixtures, and a target volume fraction. We denote the raw problem specification as
\begin{equation}
\hat{P} = \Big(S_{\mathrm{boundary}}, S_{\mathrm{force}}, \mathrm{VF}, n, a \Big),
\end{equation}
where $S_{\mathrm{boundary}}$ is a point set describing boundary-condition locations and their directional constraints, $S_{\mathrm{force}}$ is a point set describing load locations and their 2D force vectors, $\mathrm{VF}\in(0,1)$ is the target volume fraction, $n$ encodes cell/pixel size (resolution), and $a\in\mathbb{R}^2$ encodes the domain aspect ratio. Following OAT, the order-invariant point sets $(S_{\mathrm{boundary}}, S_{\mathrm{force}})$ are embedded with Boundary Point Order-invariant MLP (BPOM)-style encoders, and the remaining scalars are embedded with MLPs, producing a fixed-size conditioning vector
\begin{equation}
\begin{aligned}
P = \mathrm{concat}\Big(
&\mathrm{BPOM}_{b}(S_{\mathrm{boundary}}),\mathrm{BPOM}_{f}(S_{\mathrm{force}}),\\
&\mathrm{MLP}_{\mathrm{vf}}(\mathrm{VF}), \mathrm{MLP}_{\mathrm{cell}}(s),\mathrm{MLP}_{\mathrm{ratio}}(a)\Big).
\end{aligned}
\label{eq:oat_problem_embedding}
\end{equation}
In our later editing operators, we will occasionally need to reconstitute or update parts of $\hat{P}$ ( $\mathrm{VF}$ and the point coordinates in $S_{\mathrm{boundary}}$/$S_{\mathrm{force}}$) so that conditioning remains consistent with the edited design-domain geometry.

\paragraph{Latent Diffusion}
Let $z_0 = E(T)$ denote the clean latent encoding of a topology $T$. A forward diffusion process corrupts $z_0$ by gradually adding Gaussian noise:
\begin{equation}
z_t =\sqrt{\bar{\alpha}_t} z_0 + \sqrt{1-\bar{\alpha}_t}\epsilon,
\qquad \epsilon \sim \mathcal{N}(0,I),
\label{eq:forward_noising}
\end{equation}
where $\bar{\alpha}_t$ is a pre-defined noise schedule decreasing with timestep $t\in\{0,\dots,T\}$. OAT trains a conditional UNet to predict the velocity parameterization
\begin{equation}
v = \sqrt{\bar{\alpha}_t}\epsilon -\sqrt{1-\bar{\alpha}_t}z_0,
\end{equation}
given the noisy latent $z_t$, timestep $t$, and the problem specification $\hat{P}$. The diffusion training objective is
\begin{equation}
\mathcal{L}_{\mathrm{LDM}} = \mathbb{E}_{z_0,t,\epsilon}\Big[\big\| v - v_{\theta}(z_t, t \mid \hat{P}) \big\|_2^2\Big],
\label{eq:oat_ldm_loss}
\end{equation}
where $v_\theta(\cdot)$ is the conditional UNet predictor. 

\paragraph{Partial noising and denoising for editable sampling}
A core operation in our editor is partial noising: instead of initializing sampling from pure noise ($t=t_{total}$), we start from an intermediate noise level $t=\tau$ with $\tau \ll t_{total}$. Given an input topology $T$ we compute its latent $z_0=E(T)$ and sample a partially noised latent $z_\tau$ via Eq.~\eqref{eq:forward_noising}. We then apply an edit operator to $z_\tau$ (or to a clean reference latent that is subsequently noised) to obtain $z_\tau'$and run a deterministic DDIM-style reverse process from $\tau$ down to $0$ to obtain one or more edited latents $z_0'$, which are decoded back to density fields by the NFAE renderer.

Partial noising preserves the instance identity encoded in $z_0$ (global layout, major load-path “semantics,” and coarse structural features) because $z_\tau$ remains correlated with $z_0$ when $\bar{\alpha}_\tau$ is not too small. At the same time, injecting moderate noise introduces controllable stochastic degrees of freedom that make the representation more pliable to localized edits and allow sampling a spectrum of mechanically plausible post-edit outcomes by varying $\epsilon$. In contrast, starting from full noise would largely discard the original instance information, turning editing into full regeneration rather than an instance-preserving modification.

\subsection{Latent Edits}
\label{sec:latent_edits}

We define three edit operators that act on the latent representation of an existing topology instance. Let $z_0=E(T)$ be the clean latent code of the input density field $T$, and let $z_\tau$ denote its partially noised state obtained by applying the forward process in Eq.~\eqref{eq:forward_noising} at an intermediate noise level $\tau$. Each edit produces an edited latent initialization $z_\tau'$ and, when needed, an updated problem specification $\hat P'$ whose boundary-condition geometry is consistent with the edit. Guided denoising then maps $z_\tau' \mapsto z_0'$, and we decode $z_0'$ with the NFAE to obtain the edited topology.

\subsubsection{Topology Warp}
\label{sec:topology_warp}

\paragraph{Gaussian coordinate displacement}
A drag edit is specified by a handle location $\mathbf{h}\in[0,1]^2$, a displacement $\boldsymbol{\Delta}\in\mathbb{R}^2$, and an influence radius $\sigma>0$. We define a smooth displacement field
\begin{equation}
\mathbf{u}(\mathbf{x}) = \exp\!\left(-\frac{\|\mathbf{x}-\mathbf{h}\|_2^2}{2\sigma^2}\right)\boldsymbol{\Delta},
\qquad
\mathbf{x}\in[0,1]^2.
\label{eq:gaussian_disp_method}
\end{equation}
Using destination-to-source (inverse-map) sampling, the induced warp operator $\mathcal{W}$ acting on an image-like field $f$ is
\begin{equation}
\bigl[\mathcal{W}(f)\bigr](\mathbf{x}_{\mathrm{dst}}) =f\!\left(\mathbf{x}_{\mathrm{dst}}-\mathbf{u}(\mathbf{x}_{\mathrm{dst}})\right),
\label{eq:warp_operator_method}
\end{equation}
implemented with continuous interpolation.

\paragraph{Latent warp and BC-consistent conditioning update}
We apply the same geometric warp to the partially noised latent:
\begin{equation}
z_\tau'= \mathcal{W}(z_\tau).
\label{eq:warp_latent_init}
\end{equation}
To keep conditioning consistent with the deformed geometry, we warp the point-set components of $\hat P$ (Eq.~\eqref{eq:oat_problem_embedding}), i.e., the support and force locations. Given a source point $\mathbf{p}_{\mathrm{src}}$ (from $S_{\mathrm{boundary}}$ or $S_{\mathrm{force}}$), the corresponding destination point $\mathbf{p}_{\mathrm{dst}}$ satisfies the inverse-map relation
\begin{equation}
\mathbf{p}_{\mathrm{src}}=\mathbf{p}_{\mathrm{dst}}-\mathbf{u}(\mathbf{p}_{\mathrm{dst}}).
\label{eq:point_inverse_map_method}
\end{equation}
We solve Eq.~\eqref{eq:point_inverse_map_method} via relaxed fixed-point iteration
\begin{equation}
\mathbf{p}^{(k+1)} = (1-\rho)\mathbf{p}^{(k)} + \rho\left(\mathbf{p}_{\mathrm{src}}+\mathbf{u}(\mathbf{p}^{(k)})\right),
\qquad \mathbf{p}^{(0)}=\mathbf{p}_{\mathrm{src}},
\label{eq:fixed_point_method}
\end{equation}
with relaxation $\rho\in(0,1]$. Applying this to all points yields a warped specification $\hat P_{\mathrm{warp}}$. We use $\hat P'=\hat P_{\mathrm{warp}}$ during denoising.

\subsubsection{Lattice Infill Replacement}
\label{sec:lattice_replacement}

Lattice replacement differs from warping in that we first construct an edited reference topology in image space, then use its latent as the anchor for guided denoising.

\paragraph{Shell-infill decomposition and lattice reference}
We define a replaceable infill mask $M_{\mathrm{infill}}\in[0,1]^{H\times W}$ by identifying pixels that lie sufficiently deep inside solid regions. Let $d(p)$ be a thickness proxy evaluated for pixels $p$ inside the solid, and let $t_{\mathrm{shell}}$ be a user-specified shell thickness. We set
\begin{equation}
M_{\mathrm{infill}}(p) = \left[d(p) > t_{\mathrm{shell}}\right].
\label{eq:infill_mask_method}
\end{equation}
Given a parametric lattice pattern $L\in\{0,1\}^{H\times W}$ such as a grid or cross-shaped lattice controlled by pitch and member thickness. We form a lattice-replaced reference
\begin{equation}
T_{\mathrm{lat}} = T\odot\bigl(1-M_{\mathrm{infill}}\bigr) + L\odot M_{\mathrm{infill}}.
\label{eq:lattice_reference_method}
\end{equation}
Because $T_{\mathrm{lat}}$ changes the material volume, we recompute the induced volume fraction
\begin{equation}
\mathrm{VF}_{\mathrm{lat}} = \frac{1}{HW}\sum_{p} T_{\mathrm{lat}}(p),
\label{eq:vf_lat_method}
\end{equation}
and update the volume-fraction entry in $\hat P$ accordingly, yielding $\hat P_{\mathrm{lat}}$.

\paragraph{Latent initialization}
We encode the edited reference and use it to define the reference latent
\begin{equation}
z_{\mathrm{ref}} = E(T_{\mathrm{lat}}).
\label{eq:zref_lattice_method}
\end{equation}
We then obtain $z_\tau$ by noising $z_{\mathrm{ref}}$ using Eq.~\eqref{eq:forward_noising} at $t=\tau$, and set $z_\tau'=z_\tau$ with no additional latent-space operator is applied. During denoising we use $\hat P'=\hat P_{\mathrm{lat}}$.

\subsubsection{No-design Region Enforcement}
\label{sec:no_design_region_method}

We enforce a circular no-design region by overwriting the latent inside a prescribed disk, while allowing the diffusion model to restore the mechanical performance through diffusion.

\paragraph{Hole mask and reference latent.}
The no-design region is specified by a center $\mathbf{h}\in[0,1]^2$ and radius $\sigma>0$. Let $\mathbf{x}(p)\in[0,1]^2$ denote the normalized coordinate of pixel $p$. We define the binary hole mask
\begin{equation}
M_{\mathrm{hole}}(p)=\left[\|\mathbf{x}(p)-\mathbf{h}\|_2 \le \sigma\right],
\label{eq:hole_mask_method}
\end{equation}
and the directly holed reference topology
\begin{equation}
T_{\mathrm{hole}} = T\odot\bigl(1-M_{\mathrm{hole}}\bigr).
\label{eq:hole_reference_method}
\end{equation}
Encoding $T_{\mathrm{hole}}$ gives the reference latent
\begin{equation}
z_{\mathrm{ref}} = E(T_{\mathrm{hole}}).
\label{eq:zref_hole_method}
\end{equation}

\paragraph{Latent overwrite inside the no design region.}
We start from $z_\tau$ obtained by noising the original instance latent $z_0$ (Eq.~\eqref{eq:forward_noising} at $t=\tau$). Let $M_{\mathrm{hole}}^{z}$ denote the hole mask evaluated on the latent grid (using the same normalized $(\mathbf{h},\sigma)$). We enforce the void by overwriting the latent within the hole:
\begin{equation}
z_\tau'=z_\tau \odot \bigl(1-M_{\mathrm{hole}}^{z}\bigr)+z_{\mathrm{void}}M_{\mathrm{hole}}^{z},
\label{eq:latent_hole_overwrite_method}
\end{equation}
where $z_{\mathrm{void}}$ is a calibrated constant representing a ``void'' latent value in the autoencoder space.  We keep the boundary-condition geometry unchanged and set $\hat P'=\hat P$. 

\subsection{Guidance and Denoising}
\label{sec:guidance_denoise}

All edits share a common guided sampling procedure in the latent space. We first describe the reference guidance that biases the reverse trajectory toward a target latent, and then describe the denoising update (DDIM-style reverse step) driven by OAT’s UNet. The three edit types differ only in the choice of conditioning $\hat P'$ and reference latent $z_{\mathrm{ref}}$; among them, only topology warp additionally uses a non-constant spatial weight map to localize the effect.

\paragraph{Reference guidance.}
At a reverse-time step $t$ (starting from $t=\tau$ down to $0$), the current latent state is $z_t$. We define an $\ell_2$ reference loss on the estimated clean latent $z_0'(z_t,t)$:
\begin{equation}
\mathcal{L}_{\mathrm{ref}}(z_t)=\left\lVert z_0'(z_t,t)-z_{\mathrm{ref}} \right\rVert_2^2,
\label{eq:ref_loss_unmasked}
\end{equation}
and apply a gradient step
\begin{equation}
z_t\leftarrow z_t - s \nabla_{z_t}\mathcal{L}_{\mathrm{ref}}(z_t),
\label{eq:guidance_step_unmasked}
\end{equation}
where $s$ is a guidance step size.

\paragraph{Edit-specific choices of $z_{\mathrm{ref}}$} The reference latent for guidance helps maintain consistency between the edited and the original topology. Therefore, we select a reference latent that is close to the reference topology while still allowing the network to incorporate the edit within the region of interest. 
\begin{itemize}
\item Topology warp: we set $z_{\mathrm{ref}}=z_0$ to preserve the original instance identity away from the edit.

\item Lattice infill replacement: we set $z_{\mathrm{ref}}=E(T_{\mathrm{lat}})$, anchoring sampling to the lattice-replaced reference while allowing BC-aware structural adaptations.

\item No-design region: we define $z_{\mathrm{ref}}=E(T_{\mathrm{hole}})$, so the reference latent explicitly encodes the desired void geometry.
\end{itemize}

For topology warp, the global loss in Eq.~\eqref{eq:ref_loss_unmasked} relies on the latent $z_0$ from the reference topology. Using it directly unmasked would penalize deviation from $z_0$ at every latent location, pulling it towards the unedited reference. We therefore apply guidance only outside a neighborhood of the dragged handle via a spatial weight map on the latent grid. Let $\mathbf{h}_\star$ be the achieved handle location obtained by warping $\mathbf{h}$ with the point-warp solver in Eq.~\eqref{eq:fixed_point_method}. For latent-grid coordinate $\mathbf{x}$, define
\begin{equation}
m(\mathbf{x})=\mathrm{sigmoid}\!\left(\frac{\|\mathbf{x}-\mathbf{h}_\star\|_2-r_{\mathrm{out}}}{k}\right),
\label{eq:outer_mask_only_warp}
\end{equation}
so that $m\approx 0$ near $\mathbf{h}_\star$ (no guidance) and $m\approx 1$ far away (guidance on). The guided loss for topology warp becomes
\begin{equation}
\mathcal{L}_{\mathrm{warp}}(z_t)=\frac{\left\lVert\bigl(z_0'(z_t,t)-z_{\mathrm{ref}}\bigr)\odot m\right\rVert_2^2}{\langle m\rangle+\varepsilon},
\label{eq:ref_loss_masked_warp}
\end{equation}
and Eq.~\eqref{eq:guidance_step_unmasked} is applied with $\mathcal{L}_{\mathrm{ref}}$ replaced by $\mathcal{L}_{\mathrm{warp}}$.

\paragraph{DDIM-style denoising}
Given the guidance-adjusted latent $z_{\tau}'$, we apply a conditional DDIM reverse step driven by OAT under the conditioning of $\hat P'$. This step acts as a learned physics-aware prior that projects the edited initialization back toward the manifold of feasible topologies consistent with the updated problem specification. We estimate the clean latent by
\begin{equation}
z_0'(z_t,t)=\sqrt{\bar{\alpha}_t}z_t-\sqrt{1-\bar{\alpha}_t}v_\theta(z_t,t\mid \hat P'),
\label{eq:z0_hat_general_final}
\end{equation}
and form the corresponding noise estimate
\begin{equation}
\epsilon(z_t,t)=\sqrt{1-\bar{\alpha}_t}z_t+\sqrt{\bar{\alpha}_t}v_\theta(z_t,t\mid \hat P').
\label{eq:eps_hat_general_final}
\end{equation}
We then advance one reverse step along the deterministic DDIM trajectory:
\begin{equation}
z_{t-1}=\sqrt{\bar{\alpha}_{t-1}}z_0'(z_t,t)+\sqrt{1-\bar{\alpha}_{t-1}}\epsilon(z_t,t).
\label{eq:ddim_update_general_final}
\end{equation}
Repeating the sequence from $t=\tau$ down to $0$ yields the final edited latent, which is decoded by the NFAE renderer to produce the edited topology.

\begin{table}[t]
\footnotesize
\centering
\begin{tabular}{l|ccc}
\hline
Parameter & Warp & Lattice & No Design \\
\hline
DDIM Total Steps ($t_{total}$)   & 100  & 200 & 100 \\
DDIM Partial Steps ($\tau$)  & 20   & 20  & 25  \\
Guidance Step Size ($s$)       & 1000 & 1000 & 1000 \\
\hline
\end{tabular}
\caption{Hyperparameter configuration for the three topology editing tasks. The ratio between partial $\tau$ and total steps $t_{total}$ determines the noise level added to the latent vector. For lattice infill, we use a smaller partial-noise ratio $\tau/t_{total}$ to keep denoising in a refinement regime that preserves the reference lattice’s periodic member layout while permitting small boundary condition-aware corrections. }
\label{tab:hyperparam}
\end{table}

The hyperparameter for the three types of edit is summarized in Table \ref{tab:hyperparam}. 

\subsection{SIMP Post-processing}
\label{sec:simp_post}

We optionally refine edited designs with a short run of density-based topology optimization using the classical SIMP formulation. Here the design variable is the density field $T \in [0,1]^{H\times W}$, where $T(p)$ denotes the material density at pixel/element $p$ (1 = solid, 0 = void). Structural stiffness is interpolated from $T$ using SIMP penalization:
\begin{equation}
\mathbf{K}(T)=\sum_{p} \Big(E_{\min} + T(p)^{p}\big(E_0 - E_{\min}\big)\Big)\mathbf{K}_p,
\label{eq:simp_interpolation}
\end{equation}
where $\mathbf{K}_p$ is the element stiffness matrix for unit modulus, $E_0$ is the solid Young's modulus, $E_{\min}\ll E_0$ is a small void modulus to avoid singular stiffness, and $p=3$ is the penalization exponent that discourages intermediate densities.

In our setting, SIMP is used purely as a lightweight post-processing stage rather than a full optimization from scratch. Instead of initializing from a homogeneous field, we warm start from the diffusion-edited topology: we set the initial density to the decoded result $\hat T$ (obtained by decoding the final edited latent) and apply only a few iterations of SIMP updates using an optimality-criteria scheme under the same boundary conditions and a prescribed volume fraction $\mathrm{VF}$ (as specified in $\hat P'$). This short refinement sharpens boundaries, repairs minor local inconsistencies introduced by sampling, and improves compliance while preserving the intended edit. We apply this post-processing only for the topology warp setting, as direct lattice replacement and no design region benchmark could lead to exploding compliance due to missing boundary conditions and severing load members.

\begin{algorithm*}[t]
\caption{Latent-based topology editing }
\label{alg:topoedit}
\begin{algorithmic}[1]
\State \textbf{Encode latent:} $z_0 \gets E(T)$
\State \textbf{Add partial noise:} sample $\epsilon \sim \mathcal{N}(0,I)$ and set $z_\tau \gets q_\tau(z_0,\epsilon)$
\State Initialize $z_\tau' \gets z_\tau,\ \hat P' \gets \hat P,\ z_{\mathrm{ref}} \gets z_0$

\State \textbf{Latent edit:}
\If{\textbf{warp}}
    \State Define a displacement field $\Delta(\mathbf{x})$ from user handle/drag parameters
    \State Warp the noised latent: $z_\tau' \gets \mathcal{W}(z_\tau;\Delta)$ 
    \State Update conditioning to match the warped geometry: $\hat P' \gets \mathrm{WarpBC}(\hat P;\Delta)$

\ElsIf{\textbf{lattice infill}}
    \State Form a lattice-replaced reference topology $T_{\mathrm{lat}}$ from $(T,\ M_{\mathrm{infill}},\ L)$
    \State Update conditioning (e.g., volume fraction) to obtain $\hat P_{\mathrm{lat}}$ and set $\hat P' \gets \hat P_{\mathrm{lat}}$
    \State Convert the replaced reference to a latent target: $z_{\mathrm{ref}} \gets E(T_{\mathrm{lat}})$
    \State Re-initialize by partial noising: $z_\tau' \gets q_\tau(z_{\mathrm{ref}},\epsilon)$

\ElsIf{\textbf{no-design region}}
    \State Construct a hole mask $M_{\mathrm{hole}}$ from center $\mathbf{h}$ and radius $\sigma$
    \State Define a holed reference topology $T_{\mathrm{hole}}$ and set $z_{\mathrm{ref}} \gets E(T_{\mathrm{hole}})$
    \State Map the hole mask to the latent grid ($M_{\mathrm{hole}}^{z}$) and overwrite inside:
    \State \hspace{1.4em} $z_\tau' \gets \mathrm{OverwriteVoid}(z_\tau', M_{\mathrm{hole}}^{z}, z_{\mathrm{void}})$
\EndIf

\State \textbf{Denoising:} for $t=\tau,\dots,1$ do
\State \hspace{1.4em} Optionally apply reference guidance toward $z_{\mathrm{ref}}$ (masked/weighted for locality)
\State \hspace{1.4em} Take one DDIM reverse step using the UNet: $z_{t-1} \gets \mathrm{DDIMStep}(z_t, t, \hat P')$
\State Obtain edited latent $z_0' \gets z_{0}$ (final denoised state)

\State \textbf{Decode structure:} $\phi' \gets D(z_0')$;\quad $T' \gets R(\phi', c, s)$
\State \textbf{Optional SIMP post-processing:} run $K$ refinement steps on $T'$ to obtain $T^{\star}$
\end{algorithmic}
\end{algorithm*}

\section{Results}

In this section, we set up experiments to demonstrate the aforementioned topology warp, lattice infill replacement, and late-stage no-design region enforcement with latent-based method. For the topologies, we select from the training and testing datasets of OAT. We perform the experiment using a PC with RTX Pro 6000 graphics card with 96G of VRAM. Since the diffusion model is stochastic, we generated 64 latent edit results for each of the subsequent edit operations for the best-of-N analysis. 

\subsection{Topology Warp}

\begin{figure}
\centering

\begin{subfigure}[t]{0.25\textwidth}
\centering
\includegraphics[width=\textwidth]{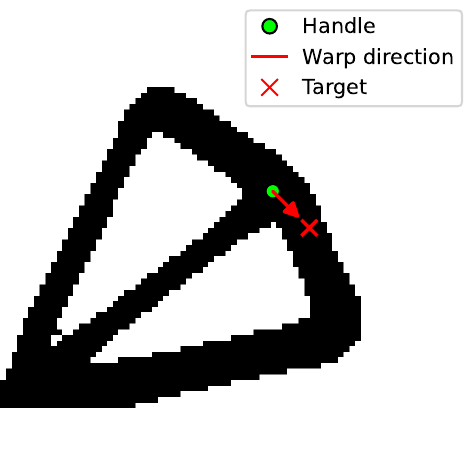}
\caption{Single warp operation}
\end{subfigure}

\begin{subfigure}[t]{0.45\textwidth}
\centering
\includegraphics[width=\textwidth]{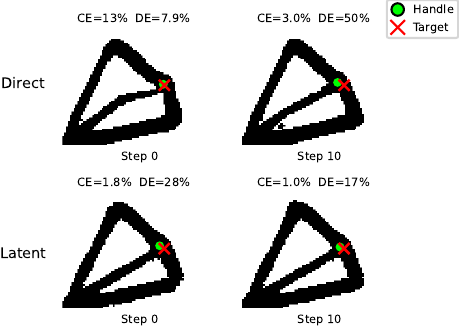}
\caption{Comparing direct and latent warp. Step 0 is no post-processing, and step 10 is after 10 steps of post-processing.}
\end{subfigure}
\caption{Illustration of a warp edit. A green handle point is selected on the topology, and a red arrow represents a warp vector. The vector’s magnitude determines the target’s location (red cross), indicating the handle point’s movement. The percentage difference in distance between the green point and the target location is calcualted Distance Error (DE). Additionally, the Compliance Error (CE) can be computed. (b) Direct warp: When the warp is performed directly on the topology, the edit causes significant local deformation, while the distance error remains relatively small. However, the compliance error increases as straight members become distorted. SIMP post-processing partially reverts some of the local changes, which reduces the distance error. Latent warp: In this case, the edit is reflected on the global topology, where post-processing has not made substantial changes to the topology. As a result, the topology remains stable, and the distance error is smaller.    }

\label{fig:warp_example}
\end{figure}

\begin{figure}
\centering

\includegraphics[width=0.45\textwidth]{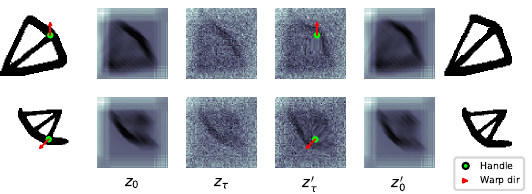}
\caption{TopoEdit workflow for warping the topology. The topology is first converted into latent $z_0$, then partial noise is added $z_\tau$, then the warp operation is mapped onto the latent space $z_\tau'$, then the warped latent is denoised to obtain the latent $z_0'$. Decoding $z_0'$ gives the edited topology. }

\label{fig:warp_latents}
\end{figure}

\begin{figure*}
\centering
\begin{subfigure}[t]{0.3\textwidth}
\centering
\caption*{  \makebox[0.33\linewidth][l]{\hspace{1em}Reference}%
  \makebox[0.34\linewidth][c]{Direct\hspace{1em}}%
  \makebox[0.33\linewidth][r]{\textbf{TopoEdit}\hspace{2em}}}
\includegraphics[width=\textwidth]{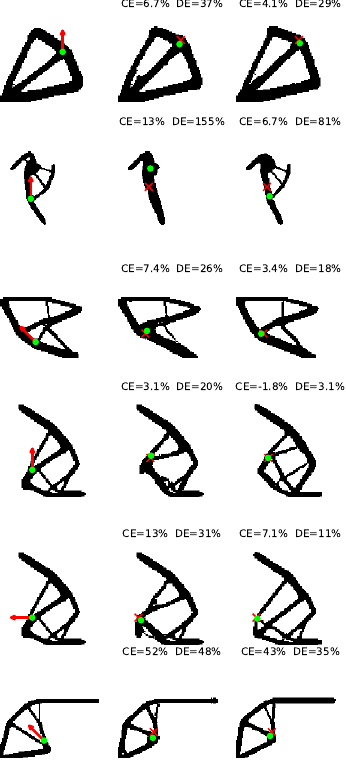}
\end{subfigure}
\qquad
\begin{subfigure}[t]{0.3\textwidth}
\centering
\caption*{  \makebox[0.33\linewidth][l]{\hspace{1em}Reference}%
  \makebox[0.34\linewidth][c]{Direct\hspace{1em}}%
  \makebox[0.33\linewidth][r]{\textbf{TopoEdit}\hspace{2em}}}
\includegraphics[width=\textwidth]{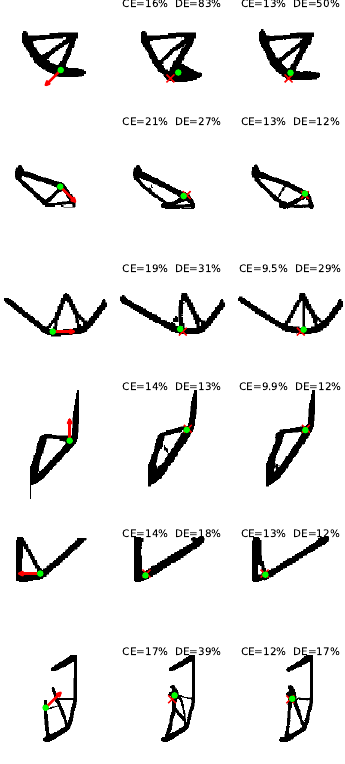}

\end{subfigure}
\qquad
\begin{subfigure}[t]{0.3\textwidth}
\centering
\caption*{  \makebox[0.33\linewidth][l]{\hspace{1em}Reference}%
  \makebox[0.34\linewidth][c]{Direct\hspace{1em}}%
  \makebox[0.33\linewidth][r]{\textbf{TopoEdit}\hspace{2em}}}
\includegraphics[width=\textwidth]{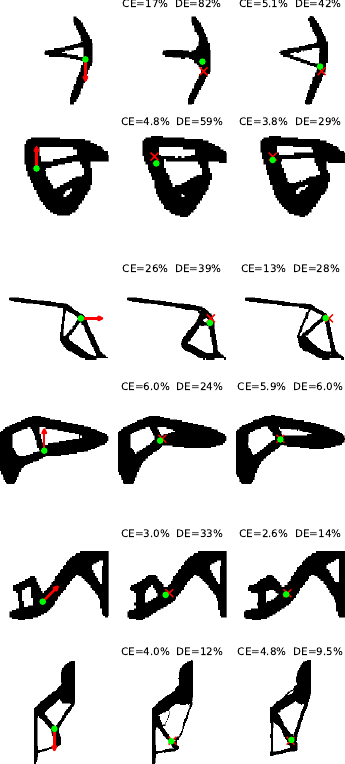}

\end{subfigure}

\caption{We select examples from the warp experiments to visualize the direct and latent-based edit. All edit results shown are after 10 steps of SIMP post-processing. The last three example is from the testing set. Left column: the reference topology. The warp edit is placed on a joint and shown as a red arrow with the end located at the point where the warp is applied and the arrow head pointing towards the target edit direction. Middle column: applying the warp edit directly to the topology, after SIMP post-processing, a larger deviation from the reference topology occurs. We show the closest joint point to the target edit location as a green dot and the edit target location as a red cross. In cases with large changes in topology after post-processing, the closest joint location is located on a different edge completely. The abrupt changes caused by the direct warp cause the post-processing optimizer to alter the topology and connectivity. Right column: with latent-based warp edits, the overall appearance of the structure is better preserved, resulting in smaller error compliance and a closer distance to the target location.    }
\label{fig:topo_warp}
\end{figure*}

\begin{figure*}
\centering
\begin{subfigure}[t]{0.12\textwidth}
\centering
\includegraphics[width=\textwidth]{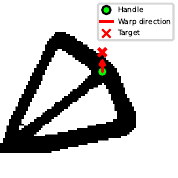}
\caption{Topology with edit}
\end{subfigure}
\qquad
\begin{subfigure}[t]{0.14\textwidth}
\centering
\includegraphics[width=\textwidth]{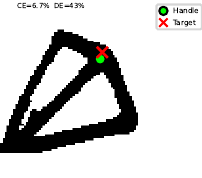}
\caption{Direct warp}
\end{subfigure}
\qquad
\begin{subfigure}[t]{0.5\textwidth}
\centering
\includegraphics[width=\textwidth]{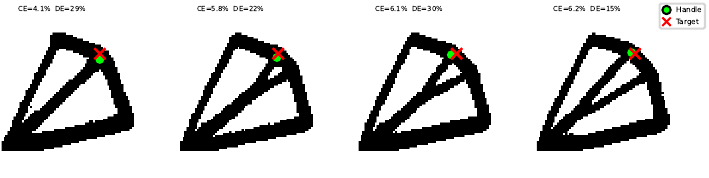}
\caption{Latent warp}
\end{subfigure}

\caption{Applying the edit (a), direct warp (b) is deterministic and produces a single result. Latent warp (c) based on the diffusion model is stochastic; the same edit on the latent can produce different results. The number of results generated can be selected as one of the configuration parameters. The stochastic nature of the diffusion model allows the user to explore a distribution of possible outcomes.  }
\label{fig:topo_warp_prob}
\end{figure*}

\begin{figure*}
\centering

\centering
\begin{subfigure}[t]{0.45\textwidth}
\centering
\includegraphics[width=\textwidth]{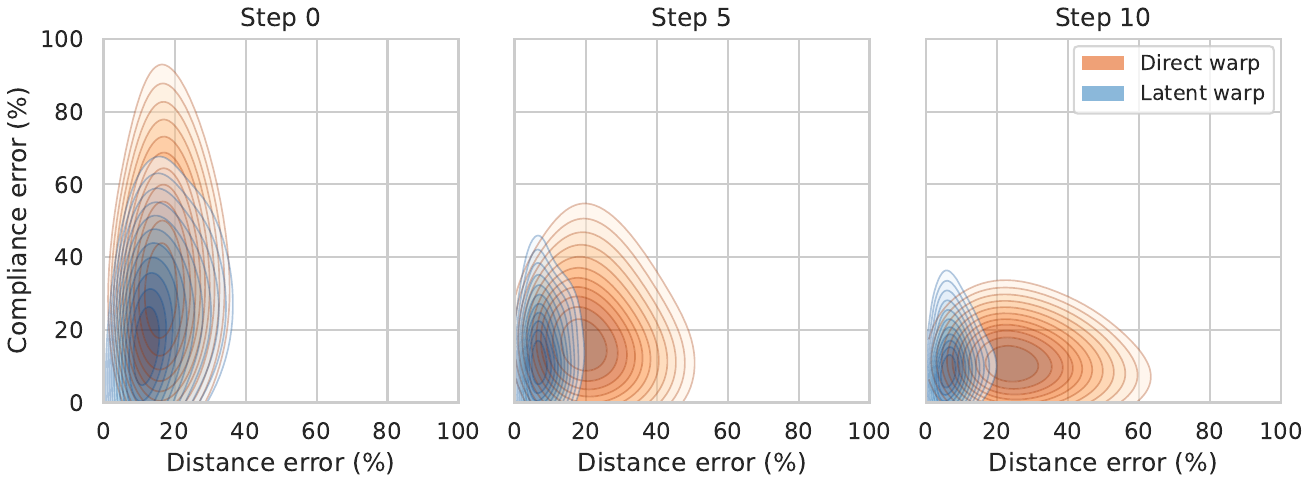}
\caption{Training set}
\end{subfigure}
\qquad
\begin{subfigure}[t]{0.45\textwidth}
\centering
\includegraphics[width=\textwidth]{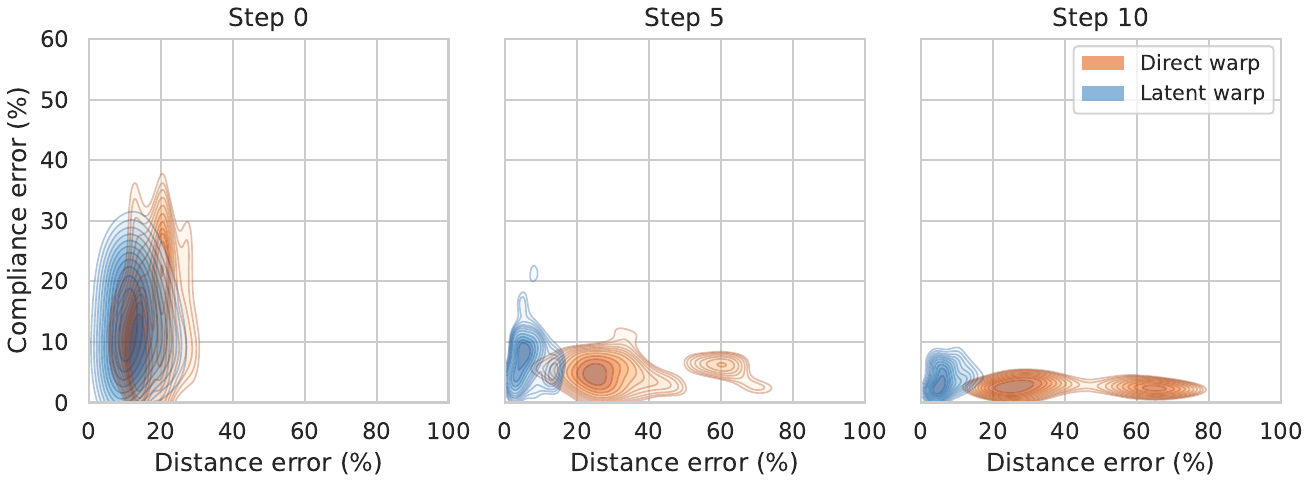}
\caption{Test set}
\end{subfigure}

\caption{We compute the distribution of the direct and latent warp (best of 64) across all warp operations for both the training and the testing set. For this plot, we apply a failure cutoff and drop cases where the compliance exceeds 1000\% of the original topology, since these cases typically correspond to broken load transfer and dominate aggregate statistics. The distribution is visualized using seaborn contours. Initially, without any post-processing (step 0), the latent warp achieved a smaller compliance error but with a slightly larger error in distance. As more steps of post-processing are applied, we observe that the center of the latent warp remains consistent, indicating that the latent warp successfully performed a more global edit. Post-processing primarily reduced the spread of the warp. In contrast, the direct warp shifted towards the right, suggesting that the local edit was reversed. }
\label{fig:warp_contour}
\end{figure*}

The OAT dataset contains entirely random boundary conditions and volume fractions. We filter the dataset based on member thickness and remove examples whose estimated thickness exceeds 10 in pixel or element units. We revisit the thicker geometries in the next subsection on lattice replacement, where replacing an interior region remains meaningful even when members are thick. After filtering, we select edit points from joints detected on the medial-axis skeletonization. This is a morphological method of finding the skeleton of an image, which is the collection of all points equidistant from the boundary. For each selected joint, we apply eight warp directions with equal angular spacing. We select a total of 122 topologies and 2208 warp operations across the filtered training set and 16 topologies and 123 warp operations across the filtered testing set. 

We focus on joint points because they provide a measurable number to evaluate edit performance. For each edited topology, we recompute the medial axis, extract skeleton junctions, and measure the distance from the intended target location to the closest detected joint. This makes it possible to quantify whether the edit moved a joint toward the desired location. For both direct warp and latent warp, we monitor compliance and the closest-joint distance across post-processing steps. We treat step 0 as the immediate output of the edit and track the metrics through the subsequent SIMP post-processing iterations. The warp operation itself, both when applied directly to the topology and when applied as a latent warp, is too fast to be reliably registered. For the latent-based edit, a single diffusion sample takes around 0.58 seconds on the GPU. We then apply 10 steps of SIMP post-processing on the CPU, which takes 1.13 seconds for both the direct and latent cases.

Figure \ref{fig:warp_example} shows a single warp operation and compares direct and latent edits across post-processing steps. For the direct warp, applying the deformation in density space causes a strong local distortion while keeping the joint displacement aligned with the intended direction, so the distance error can be small. At the same time, this local distortion bends straight members and increases compliance error. As SIMP post-processing proceeds, the optimizer tends to repair the distorted region, which reduces compliance error while also changing the achieved joint location. For the latent warp, the edit is reflected globally, and the post-processing steps introduce fewer visible changes, so the topology remains stable across steps and the achieved joint location remains close to the target. 

The mapped warp edit from the original topology onto the latent is visualized in Figure \ref{fig:warp_latents}. To highlight how this behavior appears across different geometries, Figure \ref{fig:topo_warp} visualizes selected edits after 10 steps of SIMP post-processing. In the direct warp results, post-processing often leads to larger deviations from the reference topology. In cases where the topology changes substantially during refinement, the closest joint to the target can end up on a different edge, which reflects that the optimizer altered connectivity in response to the abrupt local deformation. The latent warp results in better preservation of the overall appearance of the structure, and the achieved joint tends to remain closer to the target with smaller compliance error. 

Figure \ref{fig:topo_warp_prob} contrasts determinism and stochasticity. Direct warp on the topology is deterministic and produces a single result for a fixed edit specification, while latent warp is stochastic and the same latent warp can yield different outcomes. This stochasticity can introduce alternative local adaptations beyond a pure warp, including cases where a warped joint is interpreted as a split and members branch to form different connections. These variations give an engineer additional options to choose a desirable edit outcome, and the cost of sampling multiple candidates remains small compared to re-running full optimization from scratch.

\begin{table}[t]
\footnotesize
\centering

\begin{subtable}{\linewidth}
\centering
\caption{No post processing (Step 0)}
\begin{tabular}{l|ccc|c}
\hline
Best of & CE\% & DE\% & VFE\% & Failure Rate\% \\
\hline
2      & 28.25 & 49.09 & 1.16 & 39.06 \\
4      & 26.65 & 40.23 & 1.27 & 33.59 \\
8      & 27.23 & 36.42 & 1.25 & 25.78 \\
16     & 26.82 & 31.00 & 1.32 & 18.75 \\
32     & 24.77 & 26.25 & 1.25 & 17.19 \\
64     & 24.34 & 25.86 & 1.25 & 10.94 \\
Direct & 28.61 & 21.95 & 5.93 & 8.59 \\
\hline
\end{tabular}
\end{subtable}

\vspace{2mm}

\begin{subtable}{\linewidth}
\centering
\caption{10 steps of post processing (Step 10)}
\begin{tabular}{l|ccc|c}
\hline
Best of & CE\% & DE\% & VFE\% & Failure Rate\% \\
\hline
2      & 5.21 & 42.19 & 0.00 & 2.34 \\
4      & 4.89 & 31.97 & 0.00 & 0.00 \\
8      & 4.86 & 24.46 & 0.00 & 0.00 \\
16     & 4.76 & 19.49 & 0.00 & 0.00 \\
32     & 4.27 & 15.82 & 0.00 & 0.00 \\
64     & 4.15 & 12.27 & 0.00 & 0.00 \\
Direct & 1.76 & 50.05 & 0.00 & 0.00 \\
\hline
\end{tabular}
\end{subtable}
\caption{The best of N performance for the testing set. We see the same trend with the seaborn contour plot. The failure rate is evaluated as an edited result with more than 100\% compliance with the original structure.  }
\label{tab:warp_bon}
\end{table}

We select a representative latent result by choosing the best-of-N candidate that minimizes a sum of the compliance error and distance error. Figure \ref{fig:warp_contour} summarizes the distribution of direct and latent results across all warp operations using seaborn contours. At step 0, without any post-processing, the latent warp achieves a smaller compliance error while exhibiting a slightly larger distance error. As post-processing proceeds, the center of the latent distribution remains consistent, which indicates that the latent warp performed a more global edit and refinement mainly reduces the spread. In contrast, the direct warp distribution shifts toward larger distance error with post-processing, suggesting that the local edit is being reversed as the optimizer repairs the deformed region. Table \ref{tab:warp_bon} summarizes the best of N performance for the latent versus direct edit for the test set. We observe a similar trend with the contour plots with improving performance as more latent example sampled.

\subsection{Lattice Infill}

\begin{figure}
\centering

\includegraphics[width=0.45\textwidth]{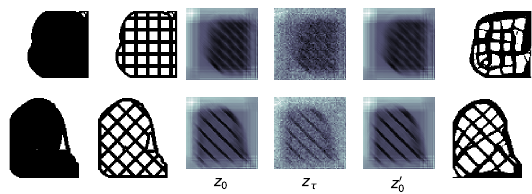}
\caption{TopoEdit workflow for lattice replacement. We take an existing topology and then apply lattice infill into it. The topology is first converted into latent $z_0$, then partial noise is added $z_\tau$, then the noised latent is denoised to obtain the latent $z_0'$. Decoding $z_0'$ gives the topology with lattice infill.}

\label{fig:lattice_latents}
\end{figure}

\begin{figure*}
\centering

\begin{subfigure}[t]{0.29\textwidth}
\caption*{Reference topology with edit overlay}
\centering
\includegraphics[width=\textwidth]{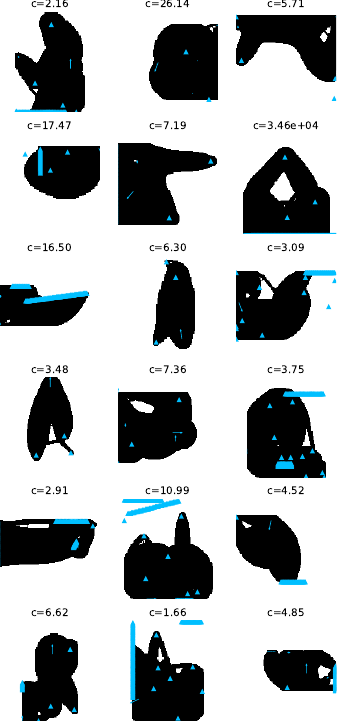}
\caption{}
\end{subfigure}
\qquad
\begin{subfigure}[t]{0.29\textwidth}
\centering
\caption*{Direct lattice replacement}
\includegraphics[width=\textwidth]{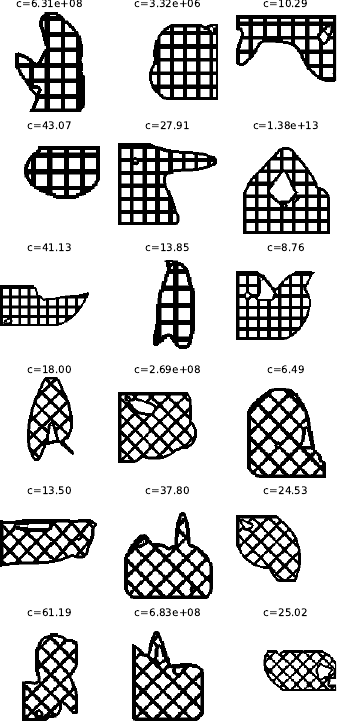}
\caption{}
\end{subfigure}
\qquad
\begin{subfigure}[t]{0.3\textwidth}
\centering
\caption*{\textbf{TopoEdit} lattice replacement}
\includegraphics[width=\textwidth]{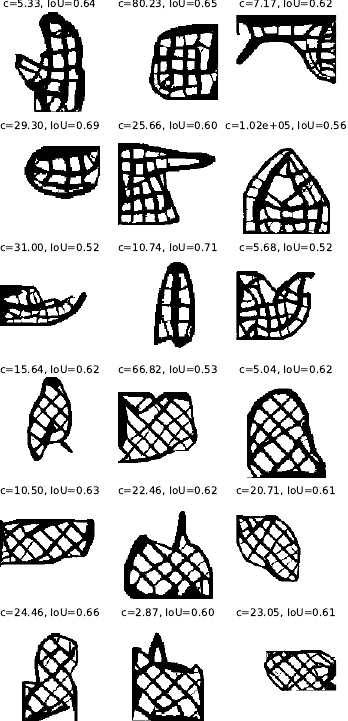}
\caption{}
\end{subfigure}
\caption{We visualize the result of direct and latent-based lattice replacement. The third and last row is from the testing set. (a) The original reference topology with boundary condition overlaid in blue and the compliance value. (b) The first three rows are using a grid-shaped lattice, and the last three rows are a cross-shaped lattice. When directly replacing the internal structure with lattices, the potential of missing a boundary condition is high, resulting in exploding compliance values. (c) With latent-based edits, the overall appearance of the lattice structure is preserved, while OAT managed to nudge the lattice structure to cover boundary condition points and align with the load paths. Resulting in lower compliance while still maintaining an acceptable IoU score to the original latent-replaced structures. }
\label{fig:lattice}
\end{figure*}

Infill/lattice control is an important design objective because it provides a route to improve manufacturability and failure modes beyond stiffness, such as buckling, enabling porous shell–infill structures that can reduce support requirements and residual-stress issues in additive manufacturing. Prior work on interactive infill topology optimization addresses this need by pausing an optimization run, prompting the user to draw an infill pattern and region of interest, and resuming optimization under an appearance constraint \cite{schiffer2024interactiveinfill}. In contrast, we treat lattice insertion as a post-optimization edit. We automatically identify replaceable interior regions using a shell–infill decomposition and employ latent diffusion denoising to adapt the inserted lattice to boundary conditions and load paths. This approach avoids catastrophic compliance failures that can occur when directly swapping infill (Figure \ref{fig:lattice} (b) cases with extremely large complaince). 

In contrast to the topology warp example, we restrict the lattice experiment to reference designs that contain sufficiently thick interior regions. We compute the medial axis of each binarized reference topology and use the associated local thickness estimate to filter examples whose interior thickness exceeds a threshold of 20 (in pixel/element units). We then run lattice replacement on 150 such reference designs for the training set and 50 for the testing set.

We evaluate mechanical performance using compliance computed under the original loads and boundary conditions, and we report it for both the direct lattice replacements and the latent reconstructions. To assess whether the latent edit preserves the intended lattice identity, we compute the Intersection over Union (IoU) between the latent result and its corresponding direct lattice replacement. We also track the volume fraction of the edited topology relative to the target design to ensure that compliance changes are not explained by increasing material usage.

We select two examples to show the partial noise and denoise with lattice replaced topology in Figure \ref{fig:lattice_latents}. While the structure still retains similar visuals to the original lattice, the structure is rearranged. Figure \ref{fig:lattice} illustrates a recurring failure mode in direct lattice swapping. When the interior region is replaced with a lattice, the resulting structure may fail to adequately cover boundary-condition points. This can lead to ill-conditioned load transfer and potentially cause compliance values to explode. In contrast, the latent-based pipeline tends to preserve the overall lattice appearance while allowing adjustments that improve mechanical performance. As shown in the examples, the diffusion model nudges lattice members to better cover boundary-condition points and to better align with dominant load paths, producing lower compliance while maintaining an acceptable IoU relative to the direct lattice-replaced reference. 

\begin{table}[t]
\footnotesize
\centering

\begin{subtable}{\linewidth}
\centering
\caption{Grid lattice replacement}
\begin{tabular}{l|cc|cc}
\hline
Best of & VFE\% & IoU & Beat Rate\% & Failure Rate\% \\
\hline
2      & 2.79 & 0.57 & 66.00 & 38.00 \\
4      & 2.58 & 0.56 & 84.00 & 30.00 \\
8      & 2.63 & 0.55 & 92.00 & 22.00 \\
16     & 2.85 & 0.55 & 98.00 & 12.00 \\
32     & 2.91 & 0.55 & 98.00 & 8.00 \\
64     & 2.47 & 0.54 & 100.00 & 8.00 \\
Direct & N/A & 1.00 & N/A & 52.00 \\
\hline
\end{tabular}
\end{subtable}

\vspace{2mm}

\begin{subtable}{\linewidth}
\centering
\caption{Cross lattice replacement}
\begin{tabular}{l|cc|cc}
\hline
Best of & VFE\% & IoU & Beat Rate\% & Failure Rate\% \\
\hline
2      & 2.10 & 0.58 & 52.00 & 46.00 \\
4      & 1.98 & 0.58 & 66.00 & 38.00 \\
8      & 1.80 & 0.58 & 76.00 & 32.00 \\
16     & 1.82 & 0.59 & 90.00 & 22.00 \\
32     & 1.70 & 0.59 & 96.00 & 16.00 \\
64     & 2.06 & 0.59 & 96.00 & 12.00 \\
Direct & N/A & 1.00 & N/A & 48.00 \\
\hline
\end{tabular}
\end{subtable}
\caption{Best of N result for the lattice replacement with the testing set. The beat rate is defined as the lattice having smaller compliance than the direct replacement. Due to the dramatic change of volume fraction and chances of missing boundary conditions, we define the failure rate here as the edited topology having 1000 times the compliance as the original structure.  }
\label{tab:lattice_bon}
\end{table}

Table \ref{tab:lattice_bon} summarizes the best of N performance on the test set for the lattice replacement experiment for the two lattice types. Across the grid- and cross-shaped lattice families, the latent lattice replacement achieves lower compliance than the corresponding direct lattice replacement. In addition, we do not observe cases where the volume fraction deviates from the target topology by more than 0.05, suggesting that the compliance gains do not come from increasing material usage. These results demonstrated the benefit of the latent edit in the lattice setting, where the diffusiom model provides a mechanism for producing lattice infills that remain recognizable while being better aligned with boundary-condition attachment and global load transfer.

\subsection{No-design Region}

\begin{figure}
\centering

\includegraphics[width=0.45\textwidth]{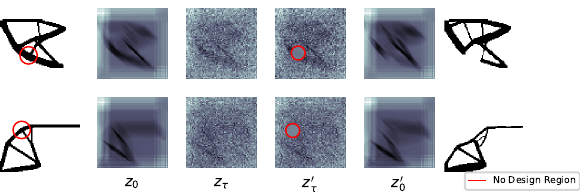}
\caption{TopoEdit workflow with no design region enforcement. The topology is first converted into latent $z_0$, then partial noise is added $z_\tau$, then the no design region is mapped as a mask onto the latent space $z_\tau'$, then the masked latent is denoised to obtain the latent $z_0'$. Decoding $z_0'$ gives the edited topology.  }

\label{fig:hole_latents}
\end{figure}

\begin{figure*}
\centering
\begin{subfigure}[t]{0.295\textwidth}
\centering
\caption*{Reference topology with edit overlay}
\includegraphics[width=\textwidth]{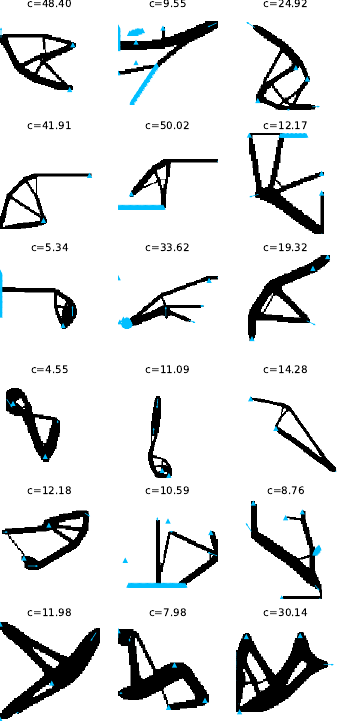}
\caption{}
\end{subfigure}
\qquad
\begin{subfigure}[t]{0.295\textwidth}
\centering
\caption*{Directly applying the no design region}
\includegraphics[width=\textwidth]{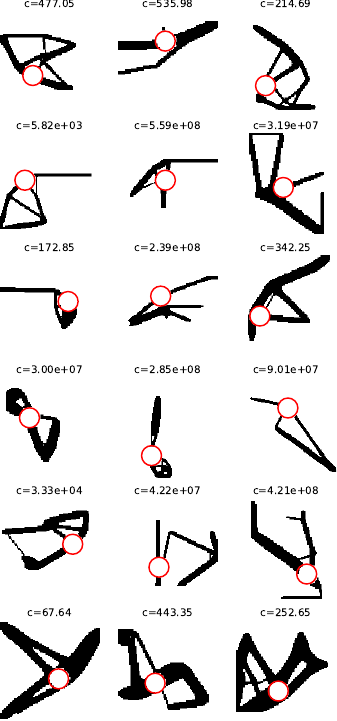}
\caption{}
\end{subfigure}
\qquad
\begin{subfigure}[t]{0.3\textwidth}
\centering
\caption*{\textbf{TopoEdit} no design region}
\includegraphics[width=\textwidth]{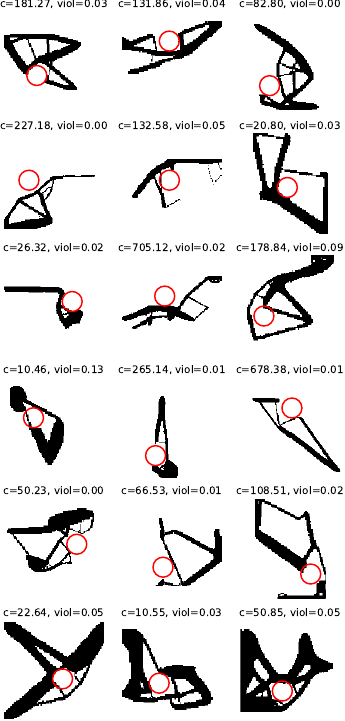}
\caption{}
\end{subfigure}

\caption{We visualize the result of directly applying the circular no design region onto the topology versus applying it in the latent vector. The last row is selected from the testing set. (a) The original reference topology with boundary condition overlaid in blue and the compliance value. (b) With the no design region directly applied to joint points, the hole masks out important load paths on the topology, causing the compliance to dramatically increase. (c) In contrast, latent-based editing managed to route around the holes and significantly reduce the compliance compared to directly applying the no design region in (b). }
\label{fig:nodesign}
\end{figure*}

No-design regions are often specified before topology optimization, since the optimization process can only account for constraints that are known at solve time. When a no-design region is introduced late in the workflow, updating an already-optimized design typically requires re-running optimization, which is time-consuming and can drift toward a different design. We therefore study late-stage no-design-region enforcement as a post-optimization edit and use it as a case study for applying a topology foundation model outside its original training distribution.

We use the same filtered subset as in the warp experiment, where the reference topologies have clearly defined joints and relatively uniform member thickness. For each reference topology, we define a circular no-design region by placing a hole at the location of a joint. We exclude candidate joints that coincide with boundary-condition points, since removing material at supports can cause the compliance to spike, and leaves little room for a feasible design to recover. In total, we evaluate 126 reference topologies for the training set and 15 topologies for the testing set. Figure \ref{fig:hole_latents} illustrates the edit operation on the noised latent $z_\tau'$ where we map the no design region on the topology as a mask onto the noised latent. 

We compare a direct baseline that applies the no-design region in density space against a latent edit that enforces the same void constraint during diffusion-based reconstruction. For evaluation, we compute compliance under the original loads and boundary conditions for both approaches. We also compute a violation ratio for the latent result, defined as the fraction of the no-design-region area that is occupied by material in the reconstructed topology, so that lower values indicate better adherence to the prescribed region.

Figure \ref{fig:nodesign} illustrates a recurring failure mode in the direct baseline. When the hole masks out an important joint and its nearby members, key load paths may be severed, which can cause the compliance to become extremely large. In contrast, the latent-based edit tends to preserve the overall structural appearance while adjusting members to route around the forbidden region, producing feasible alternatives that respect the hole while maintaining connectivity.

\begin{table}[t]
\footnotesize
\centering
\begin{tabular}{l|cc|cc}
\hline
Best of & VFE\% & Violation\% & Beat Rate\% & Failure Rate\% \\
\hline
2  & -1.07 & 4.90 & 81.25  & 12.50 \\
4  & -1.33 & 4.93 & 87.50  & 6.25  \\
8  & -0.47 & 5.08 & 93.75  & 6.25  \\
16  & -0.67 & 4.69 & 100.00 & 0.00  \\
32  & -0.63 & 4.60 & 100.00 & 0.00  \\
64  & 0.06  & 5.37 & 100.00 & 0.00  \\
Direct  & N/A  & 0.00 & N/A    & 18.75 \\
\hline
\end{tabular}
\caption{Best of N result for the no design region with the testing set. We define the beat rate as latent based no design region enforcement having smaller compliance than directly applying the no design region onto the topology. We define the failure rate as the edited topology having 1000 times the compliance as the original structure.  }
\label{tab:nodesign_bon}
\end{table}

Across all no-edit design placements, the latent-based edit achieves lower compliance than directly applying the no-design region (Table \ref{tab:nodesign_bon}). Among these results, the average violation ratio is small, indicating that the generated topologies generally avoid encroaching into the no-design region while still finding alternative load paths around the hole.

\subsection{Multiple Edits}

\begin{figure}
\centering
\includegraphics[width=0.45\textwidth]{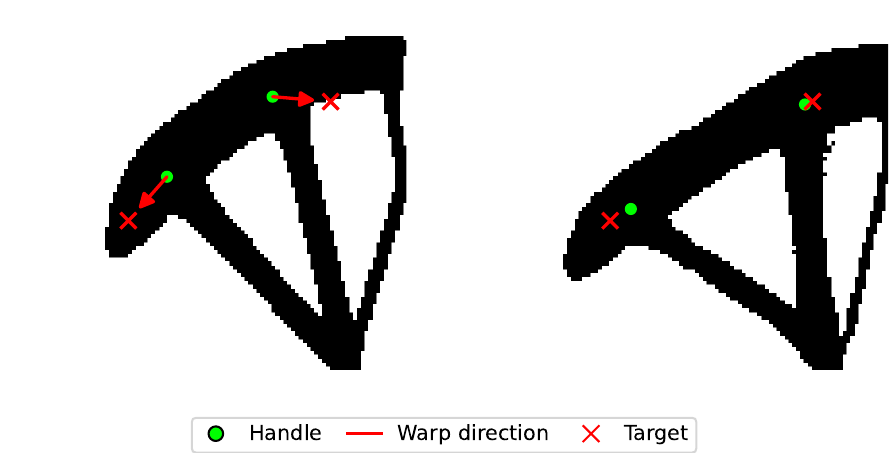}
\caption{Multiple warp operations can occur simultaneously. We apply two warp operations on the topology from the testing set (left), with the latent-based warp result shown on the right after 10 steps of post-processing. The compliance error is 26\%, and the distance error is 27\%. The higher compliance error could be due to the two edit operations causing greater changes for the topology and moving it away from the compliance optima.  }
\label{fig:two_warp}
\end{figure}

\begin{figure}
\centering
\includegraphics[width=0.45\textwidth]{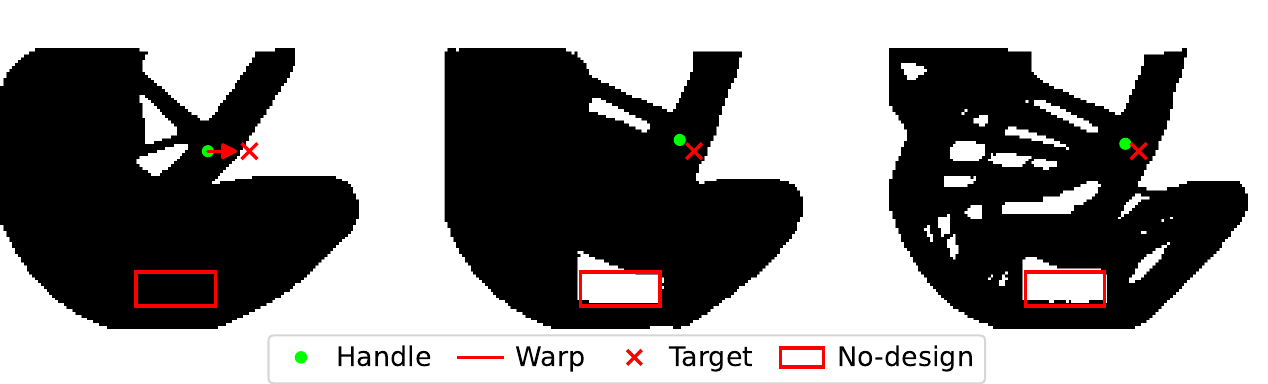}
\caption{Applying all three types of edit on a single example (left). We apply the topology warp and no design region first (middle), then apply the grid-shaped lattice infill replacement (right). The compliance of the original structure is 3.9, the middle structure is 4.12 and the lattice replaced structure is 5.74. The distance error is 43\% and 36\%. }
\label{fig:combined}
\end{figure}

In addition to the single edit demonstrated and benchmarked in the previous sections, TopoEdit supports simultaneous editing. We first showcase an example in Figure \ref{fig:two_warp} with two handle points and two different warp directions. The two handle points can be accommodated by passing in two points instead of one for the handle point $\mathbf{h}$ and the warp direction $\Delta$. With two edits, we observe a slight increase in compliance error because the two edits cause greater changes to the topology away from the compliance optima. However, the distance error remains consistent with the topology warp effect of a single handle point.

We also set up an example shown in Figure \ref{fig:combined} to demonstrate all three types of edits: topology warp, lattice replacement, and no design region combined. Due to the slightly different noise schedule of the lattice replacement compared to topology warp and no design region enforcement, we perform the warp and no design edit first. Then, we take the edited structure and apply lattice replacement. Throughout the edits, the compliance remains stable without significant increase, while the topology warp effect remains persistent. We also observe no significant violation in the no design region. 

\section{Conclusion and Future Work}
In this work, we introduced TopoEdit, a post-optimization topology editor built on a latent diffusion foundation model. By encoding an optimized topology into a structured spatial latent, applying partial noising to retain instance identity while improving editability, and performing an edit-then-denoise procedure with consistency-preserving guidance, TopoEdit enables fast, localized, and physics-aware modifications of existing designs. Across diverse case studies, latent-space editing consistently produced intention-aligned changes while better preserving mechanical performance than direct density-space edit baselines. 

While TopoEdit already demonstrates strong performance across three editing tasks, several directions can further improve both edit fidelity and robustness. First, because the editor inherits its priors from the underlying topology foundation model, advances in latent diffusion for data-driven topology optimization, with higher reconstruction accuracy, stronger conditioning generalization, and improved sampling success rates, should directly translate into improved edit performance and fewer failure cases. Second, our current editor uses a fixed-resolution latent representation; increasing latent spatial resolution and/or using richer multi-channel latents may enable finer, more localized edits.

Beyond improving the latent representation, expanding the edit operator set is another potential direction. A natural extension of the drag-based warp is explicit edge-thickening and thinning, where the user can locally increase or decrease member width while preserving connectivity and global volume constraints. More broadly, more manufacturing-aware edits can be beneficial to engineers. The lattice replacement study already suggests a surrogate route to reduce support requirements by introducing porous shell–infill structures, but an explicit treatment of overhang angle control implemented as either a latent-space constraint, a guidance term during denoising, or a learned conditioning signal would provide a more direct handle on printability and process-driven geometric feasibility.

Finally, TopoEdit can be extended to three-dimensional domains. Once a 3D latent diffusion model for topology optimization is available, the same pipeline can be generalized to volumetric warps, 3D lattice insertion, and 3D no-design enforcement, enabling post-optimization edits for practical 3D structural components while maintaining physics-aware behavior under complex boundary conditions.

\bibliographystyle{elsarticle-num}
\bibliography{References}
\end{document}